\documentclass[aps,prl,twocolumn,superscriptaddress,showpacs,floatfix]{revtex4}
\usepackage{graphicx}
\usepackage{epsfig}
\usepackage{amsmath}
\usepackage{amssymb}
\usepackage{natbib}

\begin{document}

\title{$\boldsymbol\beta$-Mn: Emergent Simplicity in a Complex Structure} 

\author{Joseph A. M. Paddison}
\affiliation{Department of Chemistry, University of Oxford, Inorganic Chemistry Laboratory, South Parks Road, Oxford OX1 3QR, U.K.}
\affiliation{ISIS Facility, Rutherford Appleton Laboratory, Chilton, Didcot, Oxfordshire OX11 0QX, U.K.}

\author{J. Ross Stewart}
\affiliation{ISIS Facility, Rutherford Appleton Laboratory, Chilton, Didcot, Oxfordshire OX11 0QX, U.K.}

\author{Pascal Manuel}
\affiliation{ISIS Facility, Rutherford Appleton Laboratory, Chilton, Didcot, Oxfordshire OX11 0QX, U.K.}

\author{Pierre Courtois}
\affiliation{Institut Laue--Langevin, 6 rue Jules Horowitz, 38042 Grenoble, France}

\author{Garry J. McIntyre}
\affiliation{Australian Nuclear Science and Technology Organisation, Lucas Heights NSW 2234, Australia}

\author{Brian D. Rainford}
\affiliation{Department of Physics and Astronomy, University of Southampton, Highfield, Southampton SO17 1BJ, U.K.}

\author{Andrew L. Goodwin}
\affiliation{Department of Chemistry, University of Oxford, Inorganic Chemistry Laboratory, South Parks Road, Oxford OX1 3QR, U.K.}

\date{\today}

\begin{abstract}

We investigate low-temperature spin correlations in the metallic frustrated magnet $\beta$-MnCo. Single-crystal polarised-neutron scattering experiments reveal the persistence of highly-structured magnetic diffuse scattering and the absence of periodic magnetic order to $T=0.05$\,K. We employ reverse Monte Carlo refinements and mean-field theory simulations to construct a simple effective Hamiltonian which accounts for the magnetic scattering. The interactions we identify describe an emergent spin structure which mimics the triangular lattice antiferromagnet. The observation of a simple collective magnetic state in a complicated crystal structure is surprising because it reverses the established paradigm of elaborate emergent states arising from many-body interactions on simple lattices. We suggest that structural complexity may provide a route to realising new states of correlated quantum matter.

\end{abstract}

\maketitle

In frustrated magnets, geometry and magnetic interactions can conspire to suppress the formation of long-range magnetic order. A theme common to the resulting disordered states is emergent behaviour: topical examples include quasiparticles resembling magnetic monopoles in spin ices \cite{Castelnovo_2008}, composite spin-orbital degrees of freedom \cite{Nakatsuji_2012}, and spinon excitations in quantum spin liquids \cite{Balents_2010}. Typically, investigations of collective behaviour have focused on insulating materials in which magnetic ions occupy a frustrated lattice such as kagome or triangular. In such cases, remarkably complex emergent states can arise from a simple geometry and set of magnetic interactions. In this Letter, we demonstrate a reversal of this paradigm: the presence of simple emergent spin structures in Co-doped $\beta$-Mn, a metal which, although chemically trivial, has a crystal structure well-known for its complexity \cite{OKeeffe_1977}. 

%However, the actual materials studied often have complex chemical compositions, reflecting a need to maximise frustration by chemical tuning \cite{Balents_2010,Mendels_2011}. In this Letter, by contrast, we identify an emergent behaviour in Co-doped $\beta$-Mn, a chemically simple but structurally complex metal \cite{Funahashi_1984}. %Neutron scattering measurements reveal that $\beta$-MnCo lacks long-range magnetic order to the lowest temperatures accessible experimentally. Nevertheless, over short length scales, peculiar magnetic interactions lead to emergent spin structures which mimic the canonical triangular lattice antiferromagnet \cite{Wannier_1950,Nakatsuji_2005}.

Pure $\beta$-Mn is a unique example of an elemental quantum spin liquid, which exhibits a host of largely unexplained magnetic phenomena, including geometrical frustration \cite{Canals_2000}, large antiferromagnetic spin fluctuations \cite{Nakamura_1997}, and non-Fermi liquid behaviour \cite{Stewart_2002}. However, all previous experimental studies of pure $\beta$-Mn have been limited by the lack of single-crystal samples. Here, we investigate two large single crystals of $\beta$-Mn, doped with 12\% and 20\% Co. The availability of single-crystal samples means that $\beta$-MnCo is a much better candidate for experiments than elemental $\beta$-Mn. Yet element and alloy are in many respects similar: the lattice parameter remains almost unchanged on Co-doping, while spin fluctuations are damped in the alloy but do not freeze out \cite{Funahashi_1984, Stewart_2009}. The $\beta$-Mn structure is primitive cubic (space group $P4_132$), with a unit cell containing 20 Mn atoms divided over two inequivalent crystallographic sites ($8c$ and $12d$) \cite{OKeeffe_1977}. Neutron diffraction reveals that Co substitutes almost entirely onto the $8c$ site \cite{Karlsen_2009}. We will show that all our data can be explained in a model in which the $8c$ site is non-magnetic, as has already been proposed for elemental $\beta$-Mn \cite{Nakamura_1997,Kohori_1993}.  Accordingly, in what follows we consider the magnetism of $\beta$-MnCo in terms of the $12d$ site only. Connections between neighbouring Mn atoms form a three-dimensional ``hyperkagome'' network of corner-sharing triangles [Fig.~1a]---a topology which has led several authors to identify the possibility of geometrical frustration \cite{Canals_2000,Nakamura_1997,Hafner_2003}.

\begin{figure}
\begin{center}
\includegraphics{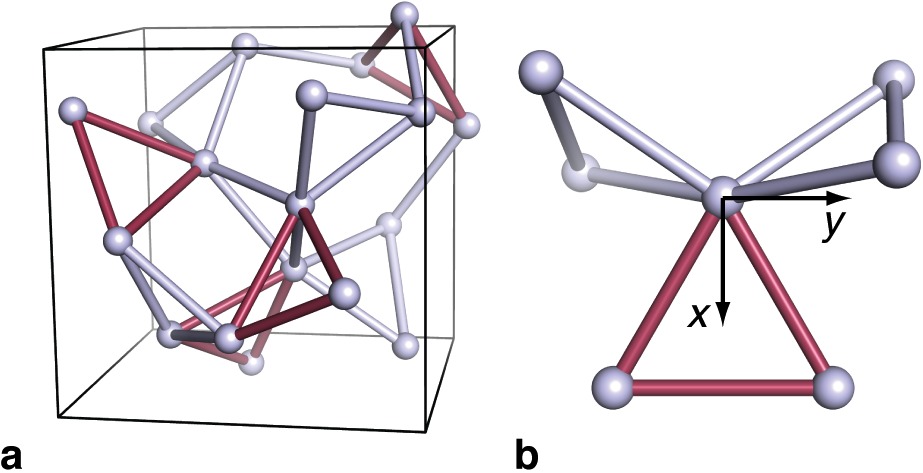}
\end{center}
\caption{\label{fig1} (a) The $\beta$-Mn crystal structure (space group $P4_132$), showing magnetic Mn atoms (12$d$ site) only. Frustrated nearest-neighbour connections are shown as grey or burgundy lines. While all connections shown are of identical length to within 1\% (see SI), only connections shown the same colour can be brought into coincidence by symmetry. (b) Local Mn environment. The six nearest neighbours of each Mn are arranged in a ``windmill'' geometry, with each triangular ``blade'' perpendicular to a $\langle111\rangle$ axis. The local $x$, $y$, and $z$ (out of the page) axes are defined for each Mn atom relative to the burgundy-coloured triangle. This choice of coordinates aligns the $x$ axis with the two-fold rotation axis which is the only point symmetry element of the 12$d$ site.}\end{figure}

Our neutron scattering experiments provide strong evidence that frustration indeed plays a central role in $\beta$-MnCo. Measurements were performed on the D7 diffractometer at the Institut Laue--Langevin, Grenoble, using $xyz$ polarisation analysis to isolate the magnetic component to the scattering \cite{Stewart_2009_2}. The sample with 20\% Co was a 17.2~g single crystal prepared using the Bridgman technique. Measurements of this sample at temperature $T=1.5$\,K are shown in four reciprocal-space planes in Fig.~2 (top left panels). These data reveal highly-structured diffuse scattering patterns---a hallmark of frustrated magnetism \cite{Bramwell_2011}. Furthermore, temperature scans reveal that this diffuse scattering remains unchanged over three orders of temperature magnitude between $T=0.05$ and $50$\,K, with no sign of the development of long-range magnetic order at $T=0.05$\,K (see SI). Importantly, measurements on the 12\% sample show that the form of the diffuse scattering is robust to Co concentration, indicating that the magnetic short-range order is unlikely to result from Co-doping, but is instead an inherent property of $\beta$-Mn (see SI). These results suggest two key questions. First, what is the microscopic origin of the persistent magnetic disorder? Second, what is the nature of the real-space spin correlations which give rise to structured diffuse scattering?

\begin{figure}
\begin{center}
\includegraphics{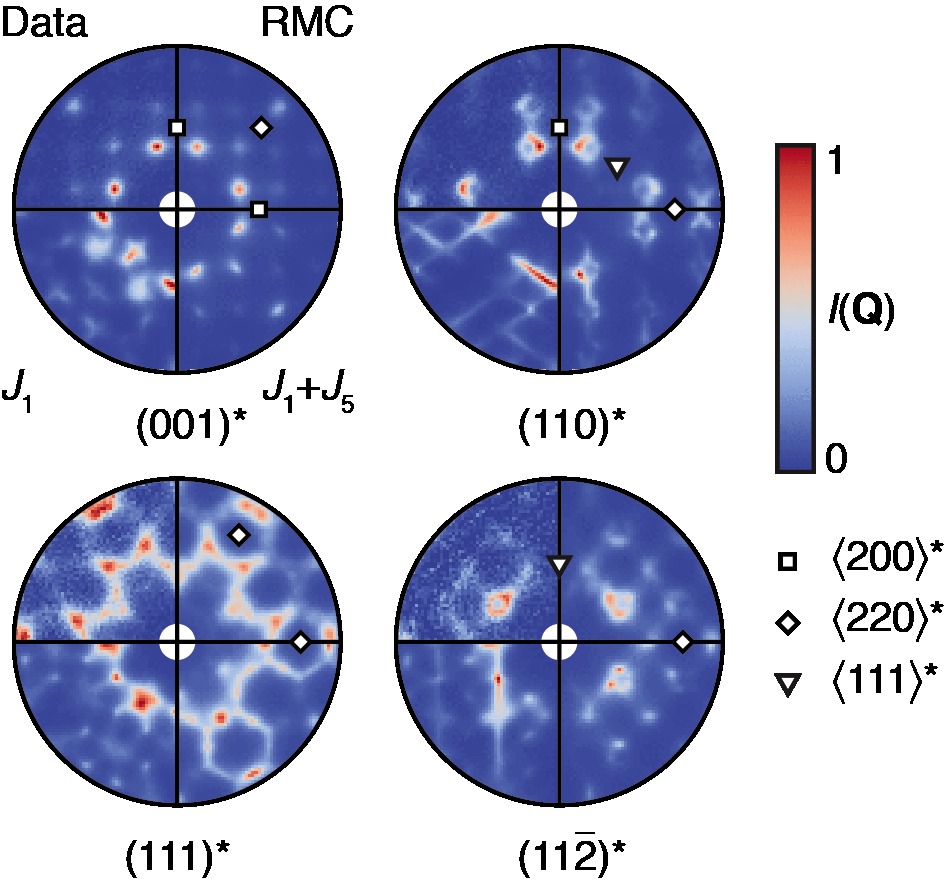}
\end{center}
\caption{\label{fig2} Magnetic neutron scattering data were collected in four reciprocal-space planes, clockwise from top left $(001)^\ast$, $(110)^\ast$, $(11\bar2)^\ast$, and $(111)^\ast$. In each plane, the top left panel shows $T=1.5\,\mathrm{K}$ experimental data, the top right panel shows the RMC fit to data, the bottom right panel shows the scattering for the $J_1$-$J_5$ model described in the text, and the bottom left panel shows the scattering for antiferromagnetic $J_1$ interactions only. Intensity scales are chosen such that the maximum value of  $I(\mathbf{Q})$ in each plane is equal to 1, and $m\bar{3}m$ Laue symmetry has been applied to experimental data and RMC fits. The $J_1$ and $J_1$-$J_5$ models are calculated using a mean-field theory at simulation temperature $T=1.01T_{\textrm c}^{\textrm{MF}}$, where $T_{\textrm c}^{\textrm{MF}}$ is the mean-field transition temperature. Whereas the $J_1$-only model is a very poor description of the data, both the RMC fit and the $J_1$-$J_5$ model give good agreement with experiment.}
\end{figure}

The simplest model which might address these points considers only antiferromagnetic interactions between nearest-neighbour spins [Fig.~1b] \cite{Canals_2000}. At the mean-field level, these interactions lead to a macroscopic degeneracy of magnetic ground states \cite{Canals_2000}. However, the diffuse scattering calculated from this model disagrees markedly with our data [Fig.~2, bottom left panels]. In order to develop a more sophisticated model, we employ a reverse Monte Carlo (RMC) approach \cite{Paddison_2012}, in which the orientations of classical spin vectors in a large configuration are refined in order to fit experimental data. The sum of squared residuals minimised during the RMC refinements is given by
\begin{equation}
\chi^{2}=\sum_{d,\mathbf{Q}}W_{d}\left[\frac{s_{d}I_{d}^{\mathrm{calc}}(\mathbf{Q})-I_{d}^{\mathrm{expt}}(\mathbf{Q})}{\sigma_{d}(\mathbf{Q})}\right]^{2},\label{eqn}
\end{equation}
where $I(\mathbf{Q})$ is the magnetic scattering intensity at reciprocal-space position $\mathbf{Q}$, superscript $\mathrm{calc}$ and $\mathrm{expt}$ denote calculated and experimental values, $s$ is a refined intensity scale factor,  $\sigma(\mathbf{Q})$ an experimental uncertainty, and $W$ an empirical weighting factor.  Subscript $d$ denotes different datasets, namely the four complete single-crystal reciprocal-space planes shown in Fig. 2 and the powder neutron scattering data of Ref. \cite{Stewart_2009}. The magnetic scattering intensity is calculated using the standard expression,
\begin{equation}
I\left(\mathbf{Q}\right)\propto\left[f\left(|\mathbf{Q}|\right)\right]^{2}\sum_{i,j}\mathbf{S}_{i}^{\perp}\cdot\mathbf{S}_{j}^{\perp}\exp\left[\mathrm{i}\mathbf{Q}\cdot\left(\mathbf{r}_{i}-\mathbf{r}_{j}\right)\right],\label{eq:sc-1}
\end{equation}
where $f(|\mathbf{Q}|)$ is the Mn\textsuperscript{4+} magnetic form factor \cite{Brown_2004}, $\mathbf{S}_{i}^{\perp}=\mathbf{S}_{i}-[(\mathbf{S}_{i}\cdot\mathbf{Q})\mathbf{Q}]/|\mathbf{Q}|^{2}$ is the component of spin $\mathbf{S}_{i}$ perpendicular to $\mathbf{Q}$, and $\mathbf{r}_{i}$ the position of spin $\mathbf{S}_{i}$.  Refinements were performed with spin configurations of size $10^{3}$ unit cells (12000 spins) using periodic boundary conditions; full details are given as SI. The fit to data obtained by our RMC refinements is shown in Fig.~2 (top right panels). In the refinements we constrain the $8c$ site to be non-magnetic---an assumption supported by the excellent agreement obtained between data and fit, but which is in apparent conflict with the conclusions of a previous RMC study \cite{Stewart_2009}. We have since identified computer-programming errors in the RMC code used in that previous study, which are entirely responsible for the misdiagnosis of a magnetic contribution from the 8\emph{c} site (see SI). In the present study, the use of single-crystal samples together with an improved RMC algorithm enables us to develop a fundamentally more robust and detailed description of spin correlations in $\beta$-MnCo. So, access to these new RMC spin configurations finally allows us to calculate the true real-space spin correlation functions responsible for the observed diffuse scattering. We consider first the radial spin correlation function $\langle\mathbf S(0)\cdot\mathbf S(r)\rangle$ shown in Fig.~3a. As anticipated from the absence of neutron scattering intensity near $\mathbf{Q}=\mathbf{0}$, strongly antiferromagnetic nearest-neighbour correlations are present. However, we find that ferromagnetic further-neighbour correlations are prominent, and note that ferromagnetic correlations between fifth-nearest-neighbours are as strong as nearest-neighbour antiferromagnetic correlations. This finding is perhaps unexpected given conventional descriptions of $\beta$-Mn as a frustrated antiferromagnet \cite{Nakamura_1997}, but may nonetheless be consistent with spin-polarised density-functional theory studies, which predict \emph{ferromagnetic} correlations within the 12\emph{d} Mn sublattice \cite{Hafner_2003}.

To complement the model-independent RMC approach---which identifies the strongest \emph{correlations}---we seek now to develop a Hamiltonian for $\beta$-MnCo that captures the key magnetic \emph{interactions}. Analysis of the RMC spin configurations indicates that the degree of spin anisotropy is small, suggesting that an effective Heisenberg model is appropriate (see SI). The Heisenberg Hamiltonian can be written
\begin{equation}
H=-\frac{1}{2}\sum_{\mathbf{R},\mathbf{\mathbf{R}^{\prime}}}\sum_{a,b}J_{ab}\left(\mathbf{R},\mathbf{R}^{\prime}\right)\mathbf{S}_{\mathbf{R},a}\cdot\mathbf{S}_{\mathbf{R}^{\prime},b},
\end{equation}
where $J_{ab}\left(\mathbf{R},\mathbf{R}^{\prime}\right)\equiv J_{n}$ is a coupling between $n$th neighbours, and each classical spin vector $\mathbf{S}_{\mathbf{R},a}\equiv \mathbf{S}_i$ is labelled by a site index $a\in\{ 1,12\}$ and a primitive lattice translation vector $\mathbf{R}$. $ $Here, the Heisenberg Hamiltonian is used as a phenomenological model, in which magnetic interactions between both localised and conduction electrons are renormalised into a small number of effective parameters. This effective local-moment approach has been widely used to study metallic systems (see, \emph{e.g.}, \cite{Wysocki_2011}). We use a mean-field theory \cite{Enjalran_2004} to calculate the paramagnetic neutron scattering pattern for a given set of $J_n$. The Fourier transforms of the $J_{ab}\left(\mathbf{R},\mathbf{R}^{\prime}\right)$ are written as a $12\times12$ Hermitian matrix $\mathbf{J}\left(\mathbf{q}\right)$, whose eigenvalues $\mathbf{\lambda}\left(\mathbf{q}\right)$ and eigenvectors $\mathbf{U}\left(\mathbf{q}\right)$ are related to the magnetic neutron scattering intensity by the expression
\begin{equation}
I_{\mathrm{MF}}(\mathbf{Q})\propto\left[f(|\mathbf{Q}|)\right]^{2}\sum_{\alpha,a,b}\frac{U_{\alpha a}(\mathbf{q})U_{\alpha b}(-\mathbf{q})}{(3-\lambda_\alpha(\mathbf{q})/T)}\exp\left[\mathrm{i}\mathbf{Q}\cdot\left(\mathbf{r}_{a}-\mathbf{r}_{b}\right)\right],\label{mf_intensity}
\end{equation}
where $\alpha,a,b\in\left\{1,12\right\} $, and $\mathbf{Q}=\mathbf{q}+\mathbf{G}$, with $\mathbf{q}$ a vector in the first Brillouin zone and $\mathbf{G}$ a primitive reciprocal lattice vector \cite{Enjalran_2004}. The interaction parameters $J_n$ were fitted to the data shown in Fig.~2 using a least-squares approach (see SI).  In order to determine the minimum number of $J_n$ sufficient to describe the data, we plot the dependence of the sum of squared residuals $\chi^{2}$ on the number of $J_n$ allowed to vary. The results of this procedure are surprising yet unambiguous [Fig.~3b]: only two interaction parameters are necessary to fit the data, with the inclusion of up to 10 further parameters leading only to marginal additional improvements in the fit. The two essential parameters are antiferromagnetic $J_1$ and ferromagnetic $J_5$, in the ratio $J_5/J_1\approx-0.65$. We note that these interactions correspond to the largest magnitudes of $\langle\mathbf S(0)\cdot\mathbf S(r)\rangle $ in Fig.~3a. The high quality of the fit to neutron scattering data obtained [Fig.~2 (bottom right panels)] is remarkable given that the model involves just two parameters.

\begin{figure}
\begin{center}
\includegraphics{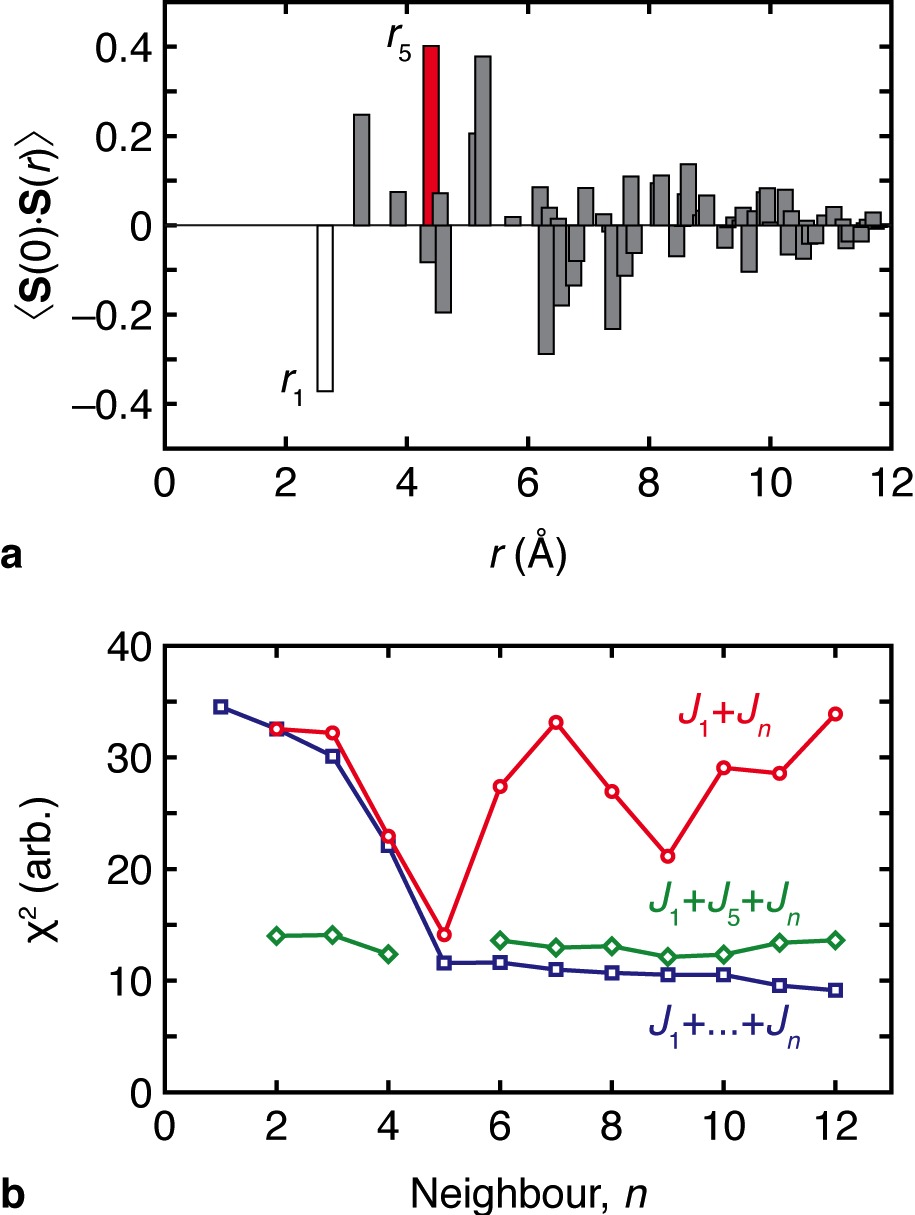}
\end{center}
\caption{\label{fig3} (a) Radial spin correlation function $\langle\mathbf S(0)\cdot\mathbf S(r)\rangle$ calculated from RMC spin configurations. (b) Variation of the sum of squared residuals $\chi^2$ as a function of the number of interaction parameters $J_n$ allowed to vary when fitting a Heisenberg model (Eq.~1) to the data. The $\chi^2$ minimum for $J_1+J_n$ (red line) at $n=5$ indicates a good fit to data, which is not significantly improved by allowing a third parameter to vary (green line) or by including all $J_n$ up to $n=12$ (blue line).}
\end{figure}

This simple $J_1$-$J_5$ Hamiltonian helps explain the absence of magnetic order in $\beta$-MnCo. In a conventional magnet, the largest
eigenvalue of the Fourier transform of the interaction matrix with elements $J_{ab}(\mathbf{R},\mathbf{R}^{\prime})$---denoted $\lambda_{\mathrm{max}}(\mathbf{q})$---has a global maximum at some wavevector $\mathbf{q}=\mathbf{q}_{\mathrm{ord}}$ in the first Brillouin zone. At the mean-field level, this is the propagation vector of the first ordered state \cite{Bertaut_1962}. By contrast, in a frustrated magnet, $\lambda_{\mathrm{max}}(\mathbf{q})$ is flat in $\mathbf{q}$, preventing the formation of an ordered state \cite{Reimers_1991}. In real materials, a completely flat spectrum of $\lambda_{\mathrm{max}}(\mathbf{q})$ is never obtained. However, the spectrum of $\lambda_{\mathrm{max}}(\mathbf{q})$ in $\beta$-MnCo is nearly flat across a large surface in $\mathbf{q}$-space (full details are given as SI), suggesting that frustration of the $J_1$-$J_5$ Hamiltonian is a key factor contributing to the lack of order. 
%\begin{figure}
%\begin{center}
%\includegraphics{fig4.jpg}
%\end{center}
%\caption{\label{fig4} Dispersion of $\lambda_{\textrm{max}}(\mathbf q)$ for the $J_1$-$J_5$ model. (a) Surface in $\mathbf q$-space for which $\lambda_{\textrm{max}}(\mathbf q)$ (defined in the text) differs by 2\% from the global maximum, $\lambda_{\textrm{max}}(\mathbf q_\textrm{ord})$. The presence of a large surface close to $\lambda_{\textrm{max}}(\mathbf q_\textrm{ord})$ indicates a large near-degeneracy of ordering wavevectors. (b) Nearly dispersionless behaviour of $\lambda_{\textrm{max}}(\mathbf q)$ in the $(hk\frac{1}{2})^\ast$ plane. The scale is chosen such that the maximum and minimum values of $\lambda_{\textrm{max}}(\mathbf q)$ in the first Brillouin zone are equal to 1 and 0 respectively. The thin horizontal line on the scale bar indicates the level at which the isosurface in (a) is plotted.}
%\end{figure}

Having obtained a credible interaction model, we proceed to determine how these interactions are manifest in the spin correlations obtained from RMC fitting. The radial spin correlation function $\langle\mathbf{S}(0)\cdot\mathbf{S}(r)\rangle $ shown in Fig.~3 does not suffice for this purpose, because the radial average obscures the relationship of the crystal structure to the spin correlations. We therefore consider the three-dimensional spin correlation function $\langle \mathbf{S}(\mathbf{0})\cdot\mathbf{S}(\mathbf{r})\rangle $. When moving from one spin to the next in this configurational average, we take account of the crystal symmetry by defining local coordinates $(x,y,z)$ which reflect the crystallographic equivalence of each Mn atom [Fig.~1]. The relationship between $J_1$ and $J_5$ interactions is shown in Fig.~4. An arbitrary central Mn atom is coupled by ferromagnetic $J_5$ interactions along helical chains [Fig.~4a]. Representing these chains as rods oriented along the helical axis maps a set of helices onto a triangular rod lattice. In this picture, the entire structure is described as four such rod lattices, oriented along the cubic $\langle111\rangle$ directions \cite{OKeeffe_1977}. The central Mn is coupled by antiferromagnetic $J_1$ to two atoms within the same rod sublattice; its four other nearest neighbours belong to the other three rod sublattices [Fig.~4c].

\begin{figure}
\begin{center}
\includegraphics{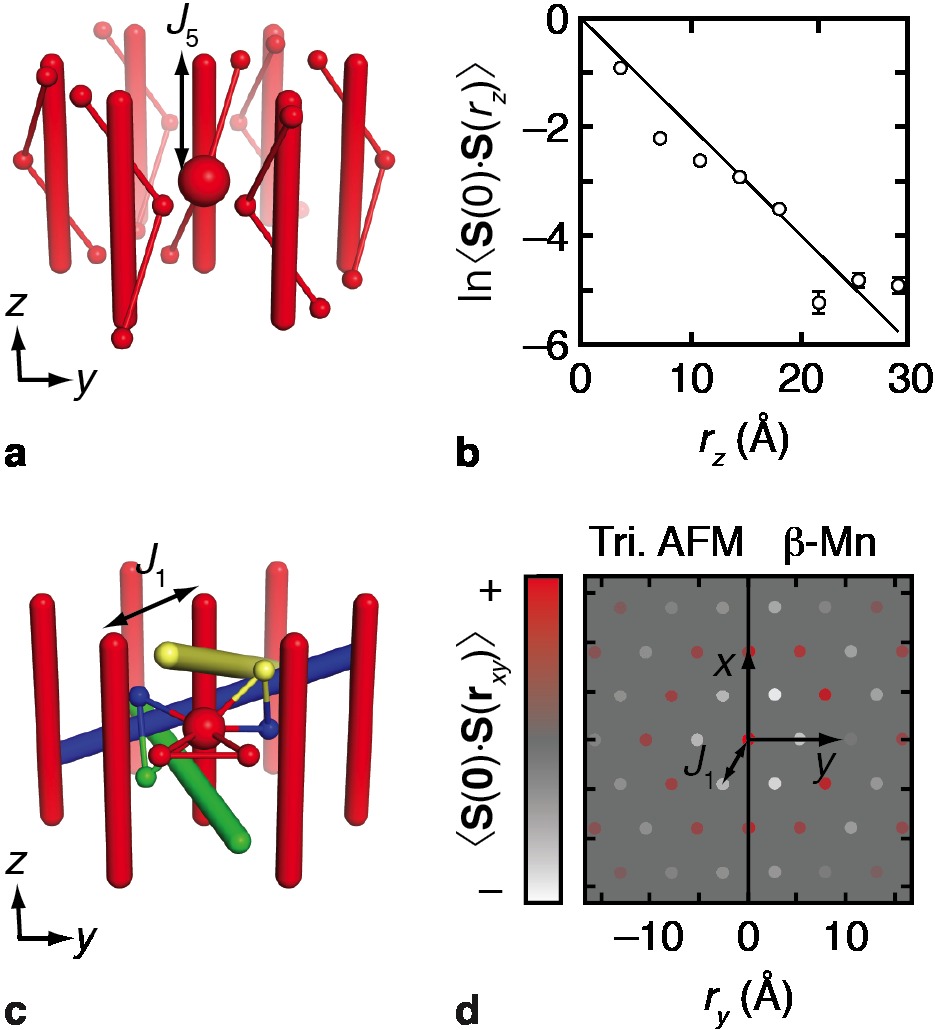}
\end{center}
\caption{\label{fig5} (a) Ferromagnetic $J_5$ interactions couple spins along helical chains, shown as thin red lines. The direction of propagation of each chain is shown as a thick red rod. The complete crystal structure can be described as four rod sublattices, each of which describes a triangular rod lattice. For clarity, only a single rod sublattice is shown. (b) Spin correlations $\langle\mathbf S(0)\cdot\mathbf S(r_z)\rangle$ along $J_5$ rods are always ferromagnetic, and decay exponentially with distance $r_z$ along the rod with correlation length $\sim$5\,\AA. (c) Antiferromagnetic $J_1$ interactions, shown as thin lines, couple spins from adjacent rods. An arbitrary central Mn atom, shown as a large red sphere, is coupled to two atoms within the same rod sublattice (coloured red); its four other nearest neighbours belong to the other three rod sublattices (coloured green, blue and yellow). (d) The right panel shows spin correlations $\langle\mathbf S(\mathbf 0)\cdot\mathbf{S}(\mathbf r_{xy})\rangle$ perpendicular to $J_5$ rods. Values shown represent a sum along each rod, as described in the text. The spin correlations alternate between antiferromagnetic and ferromagnetic throughout the triangular rod lattice. The left panel shows spin correlations $\langle\mathbf S(\mathbf 0)\cdot\mathbf S(\mathbf r)\rangle$ for a frustrated antiferromagnetic Heisenberg model on the simple triangular lattice (see SI).}
\end{figure}

The mapping of the $\beta$-Mn structure onto a rod packing suggests considering the spin correlations in two different orientations: along the $J_5$ rods, and in the plane perpendicular to them. We consider first the spin correlations along $J_5$ rods, denoted $\langle \mathbf{S}(0)\cdot\mathbf{S}(r_{z})\rangle $. This represents the correlation of our reference Mn spin with its neighbours at distance $r_{z}$ within the central rod in Fig.~4a. Values of $\langle \mathbf{S}(0)\cdot\mathbf{S}(r_{z})\rangle $ are shown in Fig.~4b: they are always ferromagnetic, with a simple exponential decay as a function of $r_{z}$. This trivial $r_{z}$ dependence means that it is possible to interpret a rod as behaving like a single collective object, and to calculate the summed spin correlation value $\sum_{r_{z}}\langle \mathbf{S}(0)\cdot\mathbf{S}(r_{z})\rangle $ for each rod within the rod packing oriented along $z$. These sums do not diverge because the ferromagnetic correlations have finite length. Values of $\sum_{r_{z}}\langle \mathbf{S}(0)\cdot\mathbf{S}(r_{z})\rangle $ are shown in Fig.~4d (right panel). The sign of $\sum_{r_{z}}\langle \mathbf{S}(0)\cdot\mathbf{S}(r_{z})\rangle $ alternates throughout the triangular rod packing, because adjacent rods are coupled by $J_1$. This pattern strongly resembles the $\langle \mathbf{S}(\mathbf{0})\cdot\mathbf{S}(\mathbf{r})\rangle $ calculated for a simple antiferromagnetic Heisenberg model on the simple triangular lattice \cite{Loison_1994, Wannier_1950}, shown in Fig.~4d (left panel). We therefore suggest that the comparable strength of $J_1$ and $J_5$ leads, over short length scales, to frustration of multi-spin rods on a triangular lattice. This emergent behaviour contrasts with the frustration of single spins on a hyperkagome lattice which would result from a $J_1$-only model \cite{Canals_2000}.

The identification of strong magnetic frustration and emergent spin structures in such a chemically-simple system provides a valuable experimental reference point against which to benchmark developments in the theory of unconventional metals \cite{Lacroix_2010,Moriya_1985}. Our results may provide a tantalising glimpse at the behaviour of pure $\beta$-Mn: the magnetic diffuse scattering is robust to Co concentration in the range for which single crystals can be prepared, while the diffuse scattering from powder samples remains essentially unchanged for the entire range of Co concentration from zero to 20\%, with only an increase in the static magnetic moment evident on doping \cite{Nakamura_1997,Stewart_2009}. But perhaps our key result is to show how typical behaviour---the emergence of complex behaviour from simple structures, as has previously been observed in metallic systems \cite{Ballou_1996,Canals_1998}---can be reversed in a structurally-complex system.  There are hints here of a more general phenomenon: the emergence of simple collective states within complex networks has also been observed in fields as diverse as neural signalling \cite{Schneidman_2006} and the dynamics of amorphous materials \cite{Ludlam_2005}. Hence a focus on structural complexity may prove an important experimental strategy for realising many of the as-yet unobserved states of correlated quantum matter \cite{Kitaev_2006}.

JAMP and ALG gratefully acknowledge financial support from the STFC, EPSRC (EP/G004528/2) and ERC (Ref: 279705). All authors are grateful to L.~C.~Chapon (ILL) for useful discussions, and to X.~Tonon, A.~R.~Wildes, P.~P.~Deen and K.~Andersen (ILL) for technical support.

%\bibliography{bmn_refs}

\begin{thebibliography}{10}

\bibitem{Castelnovo_2008}
C.~Castelnovo, R.~Moessner, S.~L. Sondhi, {\it Nature\/} {\bf 451}, 42 (2008).

\bibitem{Nakatsuji_2012}
S.~Nakatsuji, {\it et~al.\/}, {\it Science\/} {\bf 336}, 559 (2012).

\bibitem{Balents_2010}
L.~Balents, {\it Nature\/} {\bf 464}, 199 (2010).

\bibitem{OKeeffe_1977}
M.~O'Keeffe, S.~Andersson, {\it Acta Crystallogr. A\/} {\bf 33}, 914 (1977).

\bibitem{Canals_2000}
B.~Canals, C.~Lacroix, {\it Phys. Rev. B\/} {\bf 61}, 11251 (2000).

\bibitem{Nakamura_1997}
H.~Nakamura, K.~Yoshimoto, M.~Shiga, M.~Nishi, K.~Kakurai, {\it J. Phys.:
  Condens. Matter\/} {\bf 9}, 4701 (1997).

\bibitem{Stewart_2002}
J.~R. Stewart, B.~D. Rainford, R.~S. Eccleston, R.~Cywinski, {\it Phys. Rev.
  Lett.\/} {\bf 89}, 186403 (2002).

\bibitem{Funahashi_1984}
S.~Funahashi, T.~Kohara, {\it J. Appl. Phys.\/} {\bf 55}, 2048 (1984).

\bibitem{Stewart_2009}
J.~R. Stewart, R.~Cywinski, {\it J. Phys.: Condens. Matter\/} {\bf 21}, 124216
  (2009).

\bibitem{Karlsen_2009}
O.~Karlsen, {\it et~al.\/}, {\it J. Alloys Comp.\/} {\bf 476}, 9  (2009).

\bibitem{Kohori_1993}
Y.~Kohori, Y.~Noguchi, T.~Kohara, {\it J. Phys. Soc. Jpn.\/} {\bf 62}, 447
  (1993).

\bibitem{Hafner_2003}
J.~Hafner, D.~Hobbs, {\it Phys. Rev. B\/} {\bf 68}, 014408 (2003).

\bibitem{Stewart_2009_2}
J.~R. Stewart, {\it et~al.\/}, {\it J. Appl. Crystallogr.\/} {\bf 42}, 69
  (2009).

\bibitem{Bramwell_2011}
S.~T. Bramwell, {\it Introduction to Frustrated Magnetism\/} (Springer, 2011),
  chap. Neutron Scattering and Highly Frustrated Magnetism.

\bibitem{Paddison_2012}
J.~A.~M. Paddison, A.~L. Goodwin, {\it Phys. Rev. Lett.\/} {\bf 108}, 017204
  (2012).

\bibitem{Brown_2004}
P.~J. Brown, {\it International Tables for Crystallography Vol. C\/} (Kluwer
  Academic Publishers, 2004), chap. Magnetic Form Factors, pp. 454--460.

\bibitem{Wysocki_2011}
A.~L. Wysocki, K.~D. Belashchenko, V.~P. Antropov, {\it Nat. Phys.\/} {\bf 7},
  485 (2011).

\bibitem{Enjalran_2004}
M.~Enjalran, M.~J.~P. Gingras, {\it Phys. Rev. B\/} {\bf 70}, 174426 (2004).

\bibitem{Bertaut_1962}
E.~F. Bertaut, {\it J. Appl. Phys.\/} {\bf 33}, 1138 (1962).

\bibitem{Reimers_1991}
J.~N. Reimers, A.~J. Berlinsky, A.-C. Shi, {\it Phys. Rev. B\/} {\bf 43}, 865
  (1991).

\bibitem{Wannier_1950}
G.~H. Wannier, {\it Phys. Rev.\/} {\bf 79}, 357 (1950).

\bibitem{Loison_1994}
D.~Loison, H.~T. Diep, {\it Phys. Rev. B\/} {\bf 50}, 16453 (1994).

\bibitem{Lacroix_2010}
C.~Lacroix, {\it J. Phys. Soc. Jpn.\/} {\bf 79}, 011008 (2010).

\bibitem{Moriya_1985}
T.~Moriya, {\it Spin Fluctuations in Itinerant Electron Magnetism\/} (Springer,
  1985).

\bibitem{Ballou_1996}
R.~Ballou, E.~Leli\`evre-Berna, B.~F\aa{}k, {\it Phys. Rev. Lett.\/} {\bf 76},
  2125 (1996).

\bibitem{Canals_1998}
B.~Canals, C.~Lacroix, {\it Phys. Rev. Lett.\/} {\bf 80}, 2933 (1998).

\bibitem{Schneidman_2006}
E.~Schneidman, M.~J. Berry, R.~Segev, W.~Bialek, {\it Nature\/} {\bf 440}, 1007
  (2006).

\bibitem{Ludlam_2005}
J.~J. Ludlam, S.~N. Taraskin, S.~R. Elliott, D.~A. Drabold, {\it J. Phys.:
  Condens. Matter\/} {\bf 17}, L321 (2005).

\bibitem{Kitaev_2006}
A.~Kitaev, {\it Ann. Phys.\/} {\bf 321}, 2 (2006).

\end{thebibliography}
%\bibliographystyle{Science}

\end{document}